\documentclass[sigconf, authorversion, nonacm]{acmart}

\usepackage{subfigure}

\AtBeginDocument{%
  }

\begin{document}

\title{Error Mitigation in Dynamic Circuits for Hamiltonian Simulation}

\author{Sumeet Shirgure}
\email{su711773@ucf.edu}
\orcid{0009-0009-7585-1938}
\affiliation{
  \institution{University of Central Florida}
  \city{Orlando}
  \state{FL}
  \country{USA}
}

\author{Siyuan Niu}
\email{siyuan.niu@ucf.edu}
\orcid{0000-0003-4683-381X}
\affiliation{
  \institution{University of Central Florida}
  \city{Orlando}
  \state{FL}
  \country{USA}
}

\begin{abstract}
Dynamic quantum circuits integrate mid-circuit measurements and feed-forward operations to enable real-time classical processing and conditional quantum logic. These capabilities are central to key quantum protocols such as quantum error correction, and have recently demonstrated significant potential for reducing quantum resources, including circuit depth and gate count, across a range of applications. However, executing dynamic circuits on real quantum hardware introduces a critical trade-off: while resource requirements decrease, circuit fidelity degrades due to high error rates of mid-circuit measurements, as well as the decoherence errors accumulated during the extended idle periods introduced by both mid-circuit measurements and feed-forward operations.
In this paper, we systematically investigate the impact of standard error mitigation techniques on dynamic circuit applications pertaining to Hamiltonian simulation and ground state estimation of physically relevant systems like the Heisenberg model.
We explore dynamical decoupling (DD) as a strategy to suppress decoherence and crosstalk errors during idle windows introduced by mid-circuit measurements and feed-forward delays, and also examine error mitigation via zero-noise extrapolation (ZNE).
Through experiments conducted on IBM quantum hardware, we benchmark effective combinations of these strategies that maximize the practical benefits of dynamic quantum circuits in these applications.
We demonstrate that a combination of DD and ZNE is effective in mitigating the errors introduced during mid-circuit measurements and feed-forward operations, as well as the errors arising from faulty measurements. This approach yields a energy gap improvement of at least 60\% in ground state estimation and reduces observed error of time-evolved states by up to 99\% for the Ising model and up to 20\% for the Heisenberg model. 

\end{abstract}

\keywords{Error mitigation, dynamic circuits, Hamiltonian simulation}

\settopmatter{printfolios=true}

\maketitle

\section{Introduction}
\label{sec:intro}

The pursuit of practical quantum advantage in the Noisy Intermediate-Scale Quantum (NISQ) and Early Fault-Tolerant (EFT) era is largely a battle against noise and limited hardware resources. Traditionally, quantum algorithms have been viewed as static sequences of unitary gates followed by a final measurement. However, the emergence of dynamic quantum circuits, which integrate mid-circuit measurements (MCMs) and real time classical feed-forward (FF), has shifted this paradigm. By allowing the quantum processor to react to intermediate results ~\cite{carrera2024combining}, dynamic circuits enable key protocols such as quantum error correction ~\cite{brun2019quantum, roffe2019quantum}, teleportation ~\cite{gottesman1999demonstrating,postler2022demonstration}, and significantly reduced circuit depths for state preparation
~\cite{smith2023deterministic, smith2024constant, baumer2025measurement, baumer2024efficient, farrell2025digital, zi2025constant, alam2024learning, buhrman2024state,niu2024acdcautomatedcompilationdynamic,yeo2025reducing}
and algorithmic primitives ~\cite{baumer2024quantum,baumer2025approximate}.

Despite their theoretical appeal, the physical execution of dynamic circuits on current hardware introduces a critical trade-off. While they can reduce the total number of gates or depths required, the inclusion of MCMs and feed-forward operations introduces substantial classical latency. During these "waiting" periods, qubits that are not being measured or addressed remain idle, making them highly susceptible to 
 dephasing and crosstalk from neighboring measurement operations. These errors often accumulate to the point where they negate the fidelity gains achieved through reduced circuit depth.

To realize the practical benefits of dynamic circuits, it is essential to develop robust error mitigation strategies tailored to these hardware-specific overheads. In this paper, we systematically investigate the impact of two primary techniques: Dynamical Decoupling (DD) ~\cite{viola1998dynamical,souza2011robust,lidar2014review,niu2022effects}
and Zero-Noise Extrapolation (ZNE) ~\cite{temme2017error,giurgica2020digital}.
While DD is widely used to suppress idle dephasing in static circuits, its role in mitigating idling errors during the significant latencies of mid-circuit classical processing remains a frontier of study. Similarly, ZNE is broadly used to suppress errors in unitary circuits via noise amplification, but its extension to dynamic circuits requires careful consideration of how to scale noise in a regime where measurement and feed-forward errors dominate over gate errors.

We evaluate these strategies using two physically relevant models: the 1D Transverse Field Ising Model (TFIM) and the 1D Heisenberg chain. We benchmark two distinct dynamic applications: (a) ground state preparation using a constant-depth hardware-efficient ansatz, and (b) time evolution via Trotterization. Through experiments conducted on IBM quantum hardware, we demonstrate how the strategic combination of DD and ZNE can recover the fidelity of dynamic circuits when applied to quantum simulation.
In ground state estimation, we report an improvement in the energy gap of at least 60\%,
and for time evolution an improvement of $15\%-99\%$ for the Ising model and $3\%-20\%$ for the Heisenberg model.

The rest of the paper is structured as follows : Section ~\ref{sec:background} provides the necessary background on dynamic circuits, Hamiltonian simulation, and prior work.
Section ~\ref{sec:methodology} elaborates our methodology, which includes dynamical decoupling and zero noise extrapolation.
Section ~\ref{sec:experimental_setting} describes the experimental setting
of our work. Section ~\ref{sec:results} presents our results and we conclude in Section ~\ref{sec:conclusion}.

\section{Background}
\label{sec:background}

\subsection{Dynamic Circuits}
\label{sec:dynamic_circuits}
Dynamic quantum circuits extend the standard unitary model by 
integrating MCMs and real-time classical control flow within the 
qubit coherence time. Unlike static circuits, where measurements 
are set at the end and all gates are fixed at compile time, 
dynamic circuits use classical feed-forward to apply gates 
conditioned on intermediate measurement outcomes. This primitive 
enables several applications:
\begin{itemize}
    \item Error correction: stabilizer measurement and syndrome-con\-di\-tioned corrections form the core of QEC~\cite{brun2019quantum,roffe2019quantum}.
    \item Gate teleportation: joint measurements on entangled resources implement non-local gates, a building block for modular and fault-tolerant architectures~\cite{gottesman1999demonstrating,postler2022demonstration}.
    \item Depth-for-width trade-offs: measurements with conditional local gates replace deep multi-qubit unitaries, enabling const\-ant-depth long-range entanglement~\cite{baumer2024efficient,baumer2025measurement}.
\end{itemize}
Despite these advantages, executing dynamic circuits on physical hardware introduces unique noise channels. The hardware execution involves a classical control loop where measurement signals are transferred to classical electronics, processed to make a logic decision, and fed back to trigger a pulse generator. This cycle introduces hardware latency, which is often significantly longer than the duration of standard single- or two-qubit gates. During this delay, idle qubits accumulate dephasing errors. Furthermore, the measurement process itself can induce crosstalk and dephasing on spectator qubits. Addressing these hardware-specific bottlenecks through error mitigation is critical for maintaining the fidelity of dynamic operations.

\subsection{Hamiltonian Simulation}
\label{sec:hamiltonian_simulation}
The simulation of many-body systems typically involves two distinct phases: the preparation of a meaningful initial state, and the subsequent time evolution of that state.

Identifying the lowest energy configuration of a Hamiltonian is fundamental for studying condensed matter physics and quantum chemistry. In the NISQ era, this is commonly achieved through the Variational Quantum Eigensolver (VQE) ~\cite{peruzzo2014variational,tilly2022variational,cerezo2021variational}.
This hybrid algorithm utilizes a parameterized quantum circuit, or ansatz ~\cite{kandala2017hardware}, to generate a trial state $|\psi(\theta)\rangle$.
The parameters $\theta$ are iteratively optimized by a classical controller to minimize the expectation value $E(\theta)=Tr[H |\psi(\theta)\rangle\langle\psi(\theta)|]$, where $H$ is the Hamiltonian of the system.
A primary challenge in VQE is the depth-fidelity trade-off: complex Hamiltonians require expressive entangling layers that often lead to deep circuits, which are highly susceptible to noise. Using dynamic circuits for state preparation offers a unique advantage by replacing traditional entangling gates with measurement-based operations that can achieve the necessary correlations in constant circuit depth, independent of system size. 

The time evolution of quantum many-body dynamics is one of the most prominent applications of quantum computing, as the dimensionality of the Hilbert space scales exponentially with the number of particles, rendering classical numerical methods inefficient for large systems. The goal is to implement the time-evolution operator $U(t)=e^{-i H t}$.
For most physically relevant systems, such as the spin-1/2 chains considered in this work, the Hamiltonian consists of a sum of non-commuting terms, 
$H=\sum_j{H_j}$.
In digital quantum simulation, this evolution is typically approximated using Trotterization ~\cite{trotter1959product}, which decomposes the global evolution into a sequence of smaller time steps $\Delta t=t/N$

\begin{equation}
    e^{-i H t} \approx (\Pi_j e^{-i H_j \Delta t})^N
\end{equation}

This decomposition introduces an algorithmic Trotter error \cite{childs2021theory}, which can be suppressed by increasing the number of steps (i.e., taking smaller time steps). However, a practical trade-off exists: while more steps reduce the algorithmic error, they simultaneously increase the circuit depth and gate count, thereby exposing the state to greater physical hardware noise.

\subsection{Prior Work}
\label{sec:prior_work}
Recent hardware benchmarks have highlighted the significant impact of MCM induced crosstalk and classical control loop latencies on superconducting and trapped-ion processors. Efforts to suppress these errors include the development of Quantum Instrument Randomized Benchmarking (QIRB) to specifically quantify MCM error rates ~\cite{hothem2025measuring}. Furthermore, empirical studies on IBM hardware have demonstrated that tailored DD sequences can systematically outperform theoretically derived patterns when protecting qubits during the variable delays of feed-forward operations ~\cite{tong2026learning}.
A critical challenge unique to dynamic circuits is the "branching error" problem, where an incorrect mid-circuit readout leads to the execution of the wrong feed-forward gate. To counter this, researchers have proposed quasi-probabilistic error cancellation and iterative measurement error correction specifically for MCMs, showing up to a 60\% improvement in the fidelity of dynamic primitives like qubit reset and quantum teleportation ~\cite{koh2026readout}.
The community has also begun establishing holistic frameworks like "dynamarq" ~\cite{dynamarq}, designed to benchmark dynamic circuits with error mitigation, across various hardware architectures.

\section{Methodology}
\label{sec:methodology}
In this work, we use two one-dimensional spin models as benchmarks for our error mitigation strategies. Here, $n$ is the number of spins and $h$ is the relative strength of the magnetic field:
\begin{itemize}
    \begin{item} 1D Transverse Field Ising Model (TFIM): A bedrock model in statistical mechanics used to study quantum phase transitions, consisting of nearest-neighbor interactions and a transverse magnetic field.
        \begin{equation}
            H = (\sum_{i=1}^{n-1}{Z_i Z_{i+1}})+h(\sum_{i=1}^{n}{X_i})
        \end{equation}
    \end{item}
    \begin{item} 1D Heisenberg Model: A more complex model that includes $XX$, $YY$ and $ZZ$ interactions, providing a rigorous test for quantum hardware due to the high degree of non-commutativity between its terms.
        \begin{equation}
            H = (\sum_{i=1}^{n-1}{X_i X_{i+1}+Y_i Y_{i+1}+Z_i Z_{i+1}})
                +h(\sum_{i=1}^{n}{Z_i})
        \end{equation}
    \end{item}
\end{itemize}

We employ standard error mitigation techniques: dynamical decoupling (DD) and zero noise extrapolation (ZNE) in our experiments. We first introduce these methods and then explain how we adapt these methods for dynamic circuits.

\subsection{Dynamical Decoupling}
\label{sec:dd}
Dynamical Decoupling (DD) ~\cite{viola1998dynamical,niu2022effects,souza2011robust,lidar2014review}
is an open-loop quantum control technique used to suppress the interaction between a quantum system and its environment. It utilizes sequences of periodic pulses to average out unwanted environmental noise and effectively cancel the system-bath interaction.

While DD is a mature technique for static circuits, its application to dynamic circuits, especially in superconducting qubits, introduces unique challenges:
\begin{itemize}
    \begin{item}
    Measurement and Classical Processing Latency: The primary bottleneck in dynamic circuits is the "idle" time required for MCM acquisition and classical FF. These delays can be an order of magnitude longer than standard gate times. DD must be carefully scheduled to fill these specific hardware-induced gaps to protect the idle qubits from decoherence error.
    \end{item}
    \begin{item}
    Measurement Induced Dephasing: The high-power microwave pulses used for readout on one qubit can lead to crosstalk and dephasing on neighboring "spectator" qubits. Standard DD sequences must be robust enough to suppress this measurement-induced noise without the DD pulses themselves interfering with the readout signal.
    \end{item}
\end{itemize}

In this study, we evaluate how simple DD sequences can be used to mitigate the effect of these error sources and recover the energy gap of the prepared ground state and the Trotter evolution.
Specifically, we apply two $X$ pulses to the idling data qubits while the MCM/FF operations take place on the ancillas.
Since the classical processing time is not known in advance, we use IBM's 'stretch' semantic for deferred timing resolution to schedule the DD pulses.

\subsection{Zero Noise Extrapolation}
\label{sec:zne}

Zero-Noise Extrapolation (ZNE) ~\cite{temme2017error,giurgica2020digital} is a powerful error mitigation technique that estimates the noise-free expectation value of an observable by systematically amplifying the noise in a quantum circuit and extrapolating the results to the zero-noise limit. ZNE also does not require additional qubits but does require additional circuit executions.
In its standard implementation, the noise is scaled by a factor $\lambda$
(where $\lambda = 1$ is the base hardware noise), and the expectation values 
$E(\lambda)$ are fitted to a functional form. The extrapolation typically uses linear, polynomial, or exponential fits to estimate $E(0)$.

In this work, we employ "global folding" to amplify noise. For a circuit representing a unitary operation $U$, global folding scales the noise by replacing 
$U$ with $U(U^\dagger U)^k$ where the noise scaling factor is $\lambda=1+2k$.
While global folding is straightforward for static unitary gates, its application to dynamic circuits requires a more nuanced approach:

\begin{itemize}
    \begin{item}
    Unitary Dynamic Gadgets: Although our dynamic circuits involve MCMs and FF logic, they are designed to implement a deterministic unitary operation (e.g., the parity-based rotation or the constant-depth entangler as described in Section \ref{sec:experimental_setting}). For ZNE to remain valid, the entire "gadget" including the MCM and FF logic must be treated as a single unitary block.
    \end{item}
    \begin{item}
    Manual Inversion: Unlike standard gates where an automated compiler can generate an inverse, dynamic subcircuits must be inverted manually, as described in Section ~\ref{sec:experimental_setting}. This involves constructing a circuit $U^\dagger$ that undoes the conditional unitary and resets the ancillas to their original state. This manual intervention is essential for the construction of the $(U^\dagger U)$ folding sequence, to ensure the logical operation remains identity while the physical noise is amplified.
    \end{item}
    \begin{item}
    Scaling Measurements and Latency: Global folding of dynamic gadgets inherently scales not just the unitary gate errors, but also the errors associated with the MCM and the decoherence accumulated during the FF latency. This provides a more comprehensive noise scaling compared to gate-local folding, as it captures the dominant noise sources unique to dynamic logic.
    \end{item}
\end{itemize}

By applying ZNE to our manually folded dynamic circuits, we can effectively suppress the composite noise arising from both the quantum gates and the classical control loop.

\section{Experimental Setting}
\label{sec:experimental_setting}

\subsection{Ground state estimation}

\begin{figure*}
    \subfigure[]{
        \includegraphics[width=0.35\textwidth]{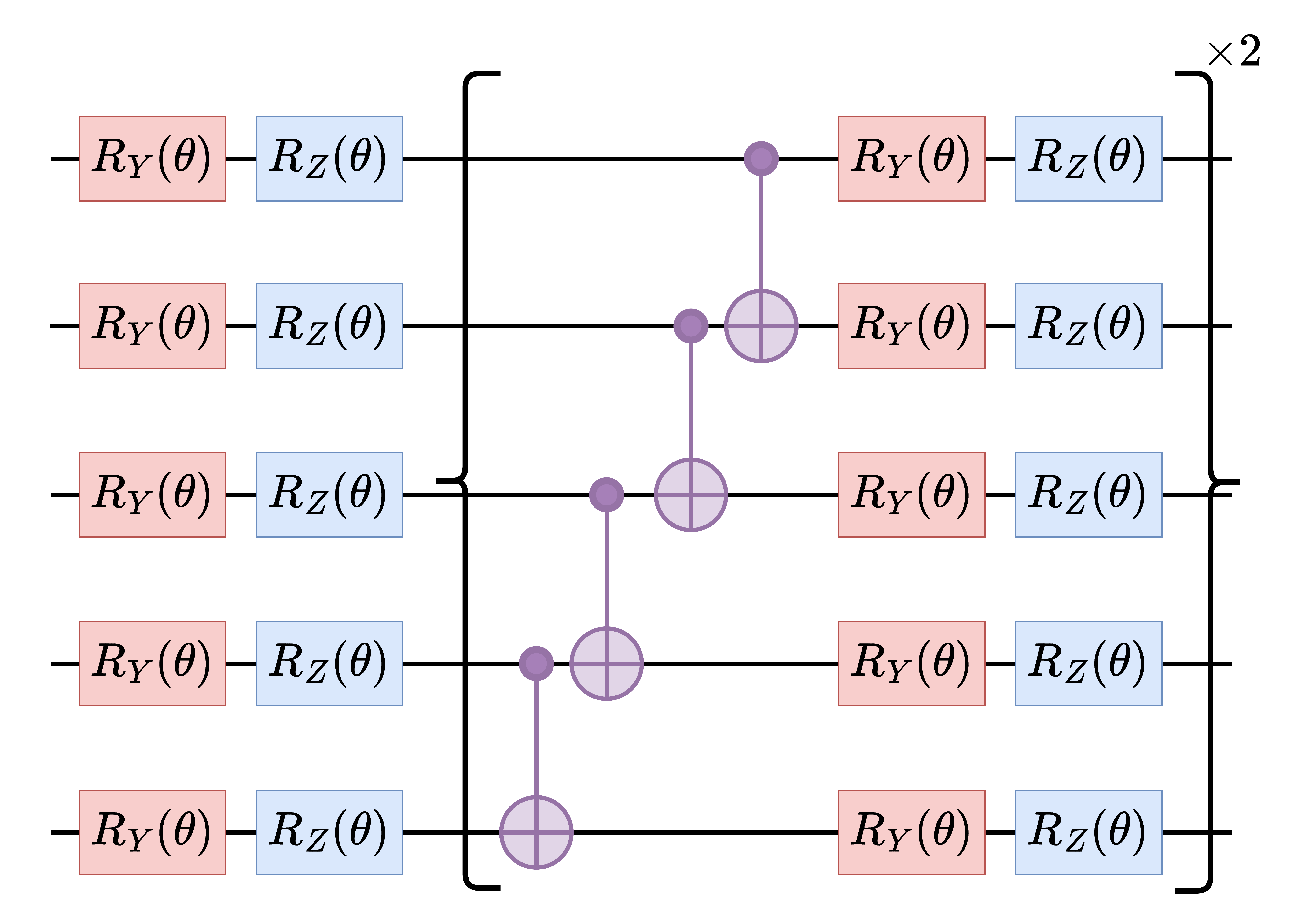}
        \label{fig:hea}
    }
    \hspace{30pt}
    \subfigure[]{
        \includegraphics[width=0.37\textwidth]{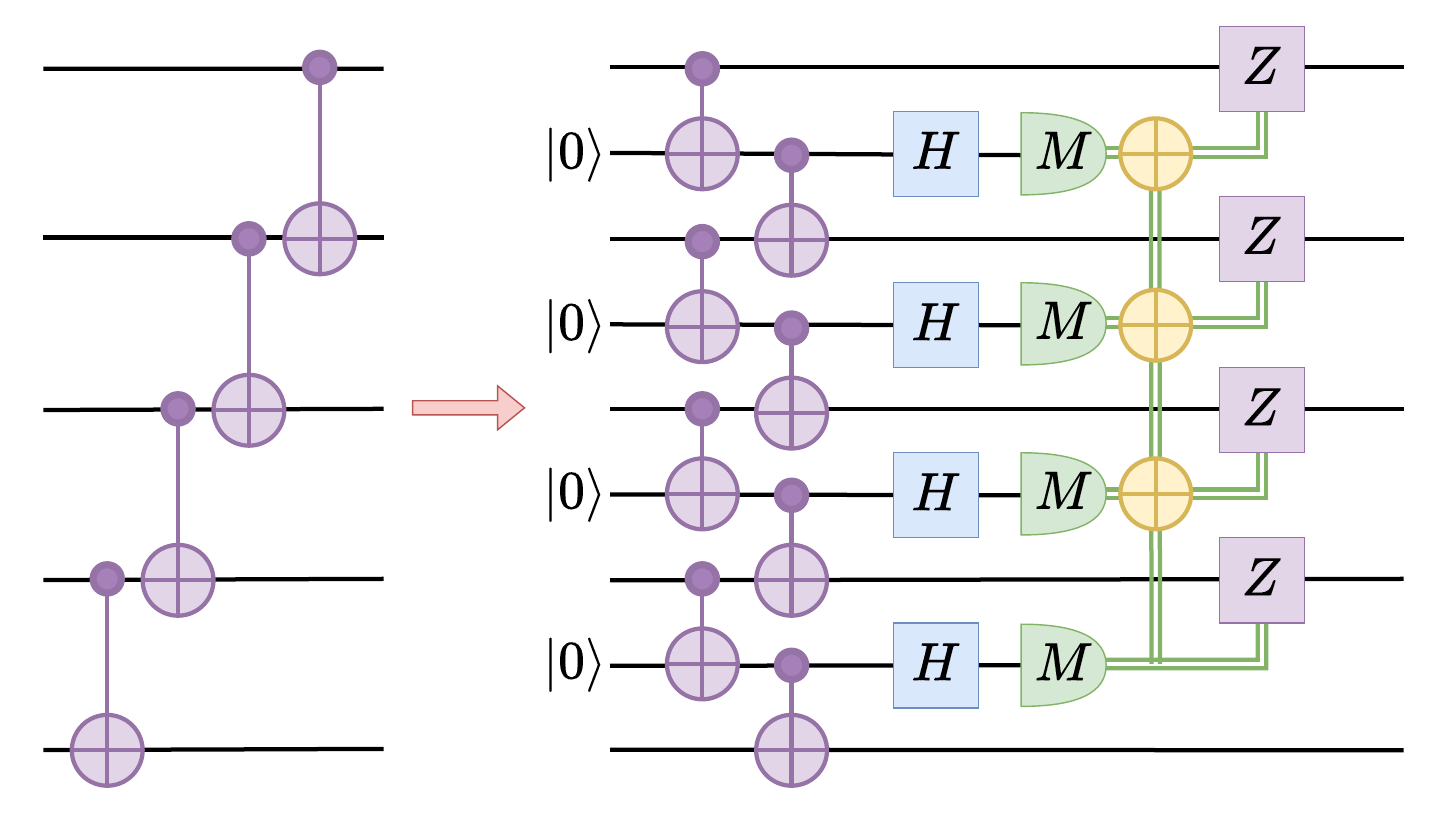}
        \label{fig:entangler}
    }
    \caption{
    (a) Hardware efficient ansatz with two entangling layers. Each Pauli rotation has a dedicated parameter $\theta_j$.
    (b) A constant depth dynamic circuit implementing the entangler.
    }
\end{figure*}

The first phase of our study evaluates the efficacy of dynamic circuits and error mitigation in preparing the ground states of the 1D Transverse Field Ising Model and the 1D Heisenberg model.

We employ a two-layer hardware-efficient ansatz (HEA) ~\cite{kandala2017hardware},
depicted in Figure ~\ref{fig:hea}.
To leverage the advantages of dynamic circuits, we replace the standard linear CNOT entangling ladder with a dynamic entangler based on the protocol introduced in ~\cite{baumer2025measurement}. This dynamic circuit is shown in Figure ~\ref{fig:entangler}.
This implementation allows for the generation of the required entanglement structure in constant circuit depth by utilizing MCMs and FFs, significantly reducing the gate-depth overhead as the system size increases.

\begin{figure}
\centering
    \includegraphics[width=0.35\textwidth]{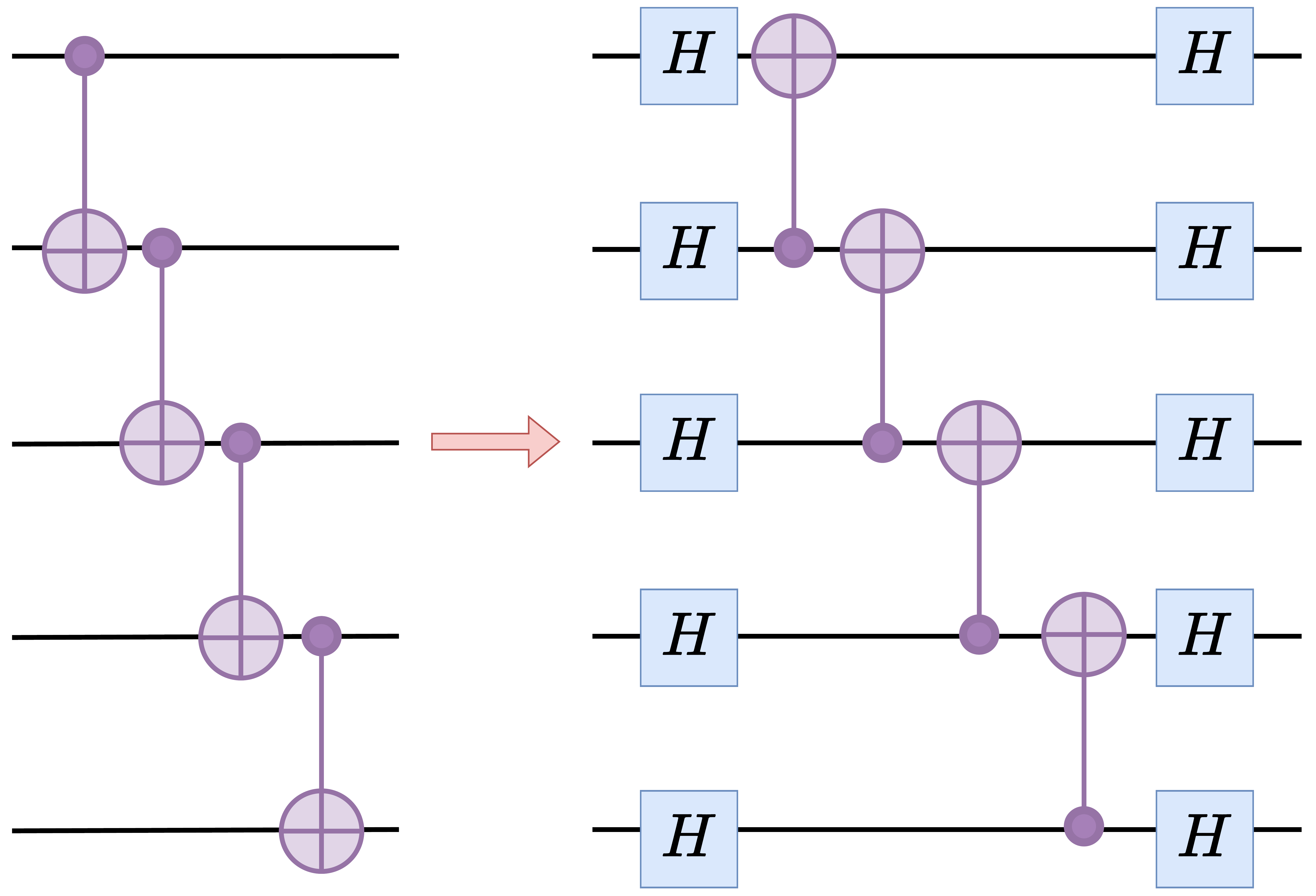}
    \caption{Inverse of the entangler needed in the hardware-efficient ansatz, which is required for folding circuits for ZNE.}
    \label{fig:inverse_entangler}
\end{figure}

For ZNE folding, the inverse of the CNOT ladder is required.
It can be realized by constructing the same ladder but with the direction reversed, followed by a conjugation with Hadamards. Figure ~\ref{fig:inverse_entangler} shows this construction.

To ensure the hardware benchmarks reflect physical noise rather than optimization variance, we determine the variational parameters via classical simulation for each combination of number of spins $n\in\{3,5,8,12\}$ and field strength $h\in\{0,0.5,2,5\}$.
Given the high cost of hardware based training and the fact that hardware noise degrades the signal given to the classical optimizer, this "best-case" parameter set allows us to focus exclusively on the hardware's ability to execute the state preparation circuit.
We denote the energy of this classically optimized wavefunction as $E_{ideal}$, which serves as our theoretical reference point.
The optimized circuits are executed on IBM Kingston backend across six distinct configurations to isolate the benefits of each mitigation strategy:
\begin{itemize}
    \item Baseline ($E_{baseline}$): 0-fold (standard) execution without DD.
    \item ZNE-only ($E_{ZNE(0,1,2)}$): 1-fold and 2-fold global folding without DD.
    \item DD-only ($E_{DD}$): 0-fold execution with DD sequences applied during measurement and feed-forward latencies.
    \item Integrated ($E_{DD+ZNE(0,1,2)}$): 1-fold and 2-fold global folding with DD.
\end{itemize}

To quantify performance, we define the baseline energy gap
$\Delta E_{base}=|E_{baseline}-E_{ideal}|$.
The relative improvement for each strategy is then calculated as the percentage reduction in this energy gap:
\begin{itemize}
    \item Improvement from DD :
    $(1-\frac{|E_{DD}-E_{ideal}|}{\Delta E_{base}})\times100\%$
    \item Improvement from ZNE :
    $(1-\frac{|E_{ZNE(0,1,2)}-E_{ideal}|}{\Delta E_{base}})\times100\%$
    \item Combined improvement :
    $(1-\frac{|E_{DD+ZNE(0,1,2)}-E_{ideal}|}{\Delta E_{base}})\times100\%$
\end{itemize}

This tiered approach allows us to determine if DD and ZNE provide additive benefits or if one technique dominates the mitigation of errors in dynamic state preparation.

\subsection{Time evolution}

\begin{figure*}
    \subfigure[]{
        \includegraphics[width=0.50\textwidth]{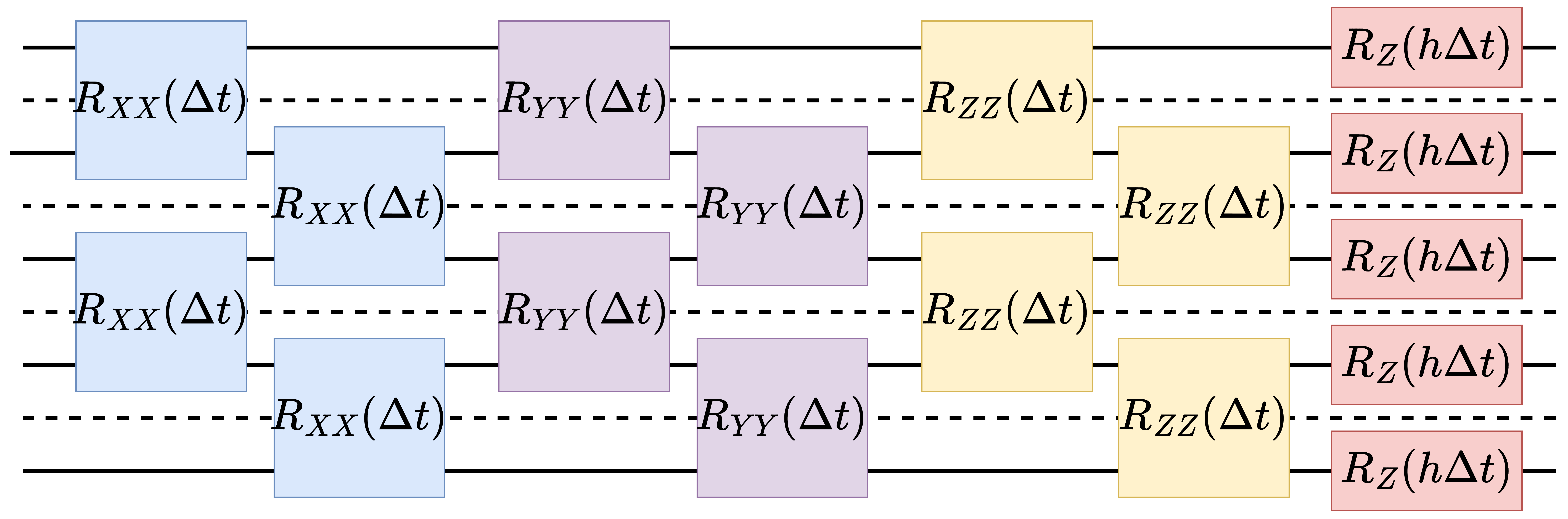}
        \label{fig:trotter_step}
    }
    \hspace{30pt}
    \subfigure[]{
        \includegraphics[width=0.40\textwidth]{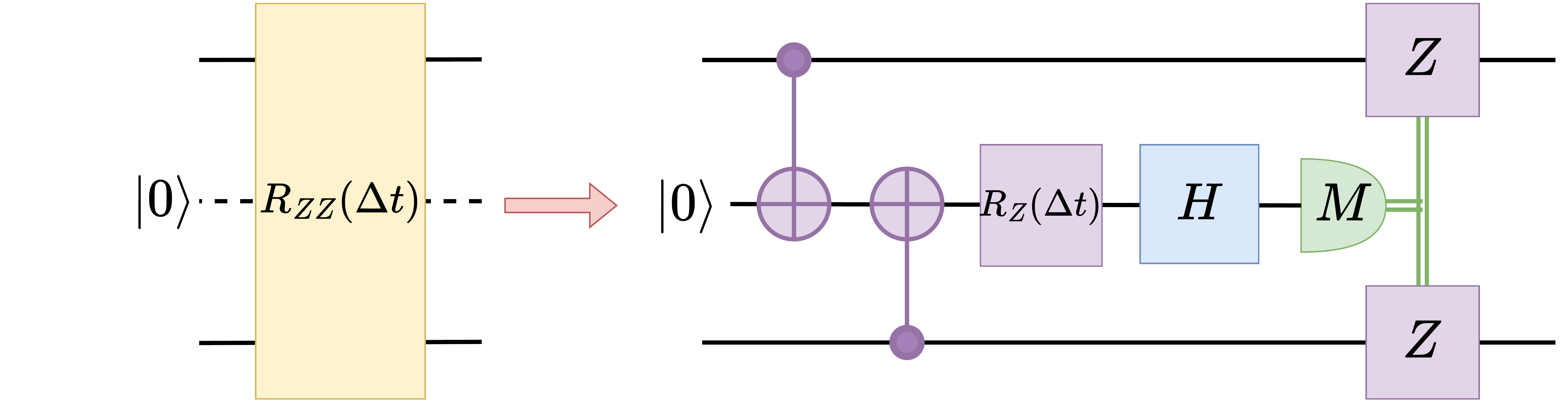}
        \label{fig:dynamic_RZZ}
    }
    \caption{
    ~\ref{fig:trotter_step} A "brickwork" circuit for a Trotter step in Heisenberg Hamiltonian simulation. The structure for the Ising Hamiltonian is similar - replace $R_Z$ gates with $R_X$ gates and only use $R_{ZZ}$ rotations. The dotted lines indicate ancillas that are reset before every instantiation of the gadget.
    ~\ref{fig:dynamic_RZZ} Ancilla mediated dynamic circuit gadget implementing an
    $R_{ZZ}(\Delta t)$ gate.
    }
\end{figure*}

We investigate the performance of dynamic circuits in out of equilibrium scenarios, by simulating the time evolution of the 1D TFIM and Heisenberg models.
To separate the errors inherent in Trotterized evolution from those introduced during
state preparation, we initialize the system in an all zero state $|0\rangle^{\otimes n}$.
This choice provides a high fidelity starting point, ensuring that the observed degradation in energy is primarily attributable to the gates and measurement logic cycles within the dynamics. We perform these simulations for system size $n\in\{5,12\}$ and $h\in\{0.5,5\}$.
We set $t=0.25$ and $N$(number of Trotter steps)$=5$. Note that system size is number of qubits without MCMs.

We implement the time-evolution operator using a first-order Trotter decomposition, depicted in Figure ~\ref{fig:trotter_step}.
To optimize execution on hardware, we arrange the interactions in a two-layer "brickwork" circuit architecture, which allows for maximal parallelism of the gates.

Each $R_{ZZ}(\Delta t)$ rotation is implemented using the parity-based ancilla gadget shown in Figure ~\ref{fig:dynamic_RZZ}.
This involves entangling a pair of data qubits with an ancilla, applying a rotation on the ancilla, performing a mid-circuit measurement of the ancilla, followed by a feed-forward correction. For the Heisenberg model, the $R_{XX}(\Delta t)$ and $R_{YY}(\Delta t)$ terms are implemented by conjugating the $R_{ZZ}(\Delta t)$ dynamic gadget with appropriate local Clifford basis-change gates.
To minimize noise and latency, and to make the DD sequences align, all MCMs and FFs
in one layer of the brickwork circuit are executed simultaneously.

For the ZNE component of the experiment, we apply global folding to the dynamic circuits. This again requires the manual implementation of the inverse operation.
For the two qubit rotations, the inverse is realized as $R_{ZZ}(-\Delta t)$, utilizing the same dynamic gadget structure but with the negated ancilla rotation angle.
This ensures that the $(U^\dagger U)$ folding sequence captures the full noise profile of the measurement and feed-forward cycle.

In a closed quantum system, the total energy $\langle H \rangle$ is a conserved quantity under the evolution of $e^{-i H t}$.
However, hardware noise including dephasing during classical latencies and measurement errors, causes the observed energy to drift from its initial value. Because noise can stochastically increase or decrease the expectation value, resulting in high variance, we utilize the same percentage reduction in the energy gap (the deviation from the initial energy) to evaluate the efficacy of DD, ZNE, and their combination.

\section{Results}
\label{sec:results}

\begin{figure*}
\centering
    \includegraphics[width=0.9\textwidth]{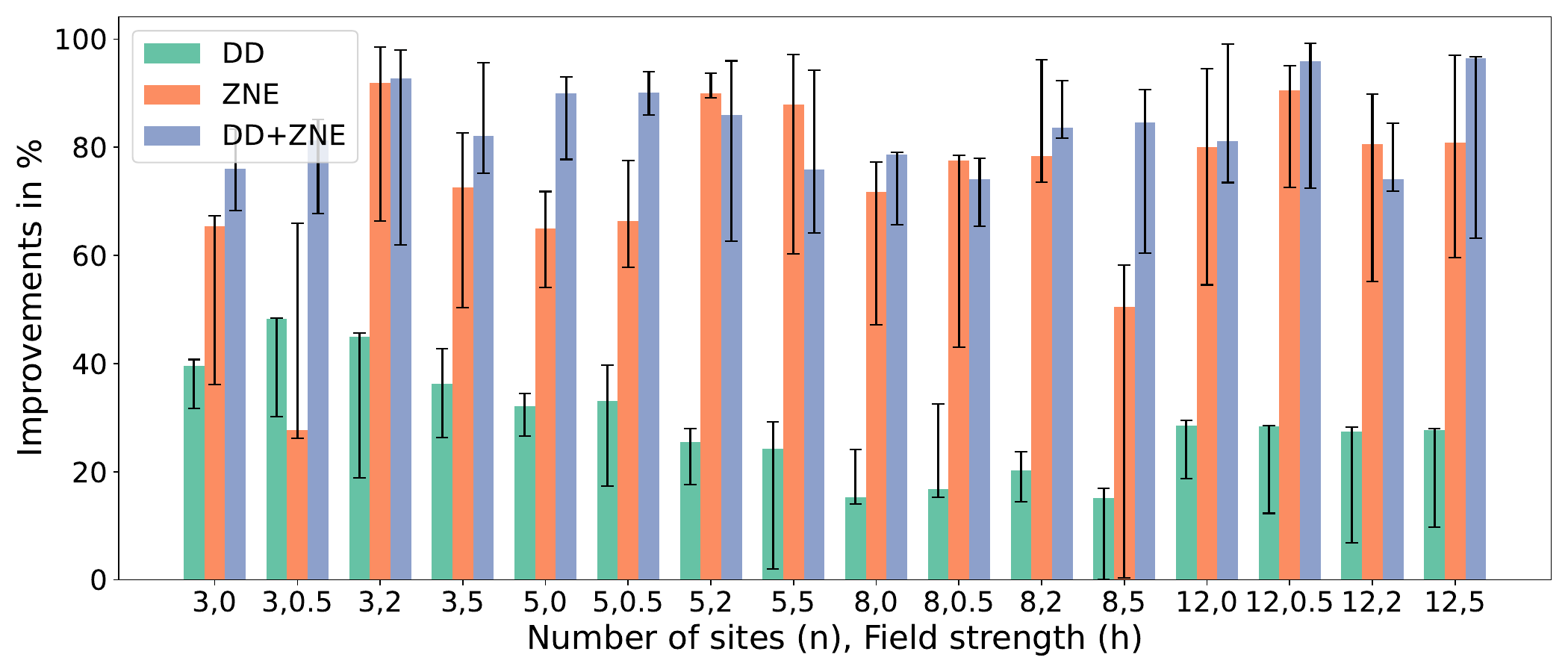}
    \caption{Percentage improvements in energy gap for Heisenberg model ground state estimation.}
    \label{fig:heisenberg_gsp}
\end{figure*}

\begin{figure*}
\centering
    \includegraphics[width=0.9\textwidth]{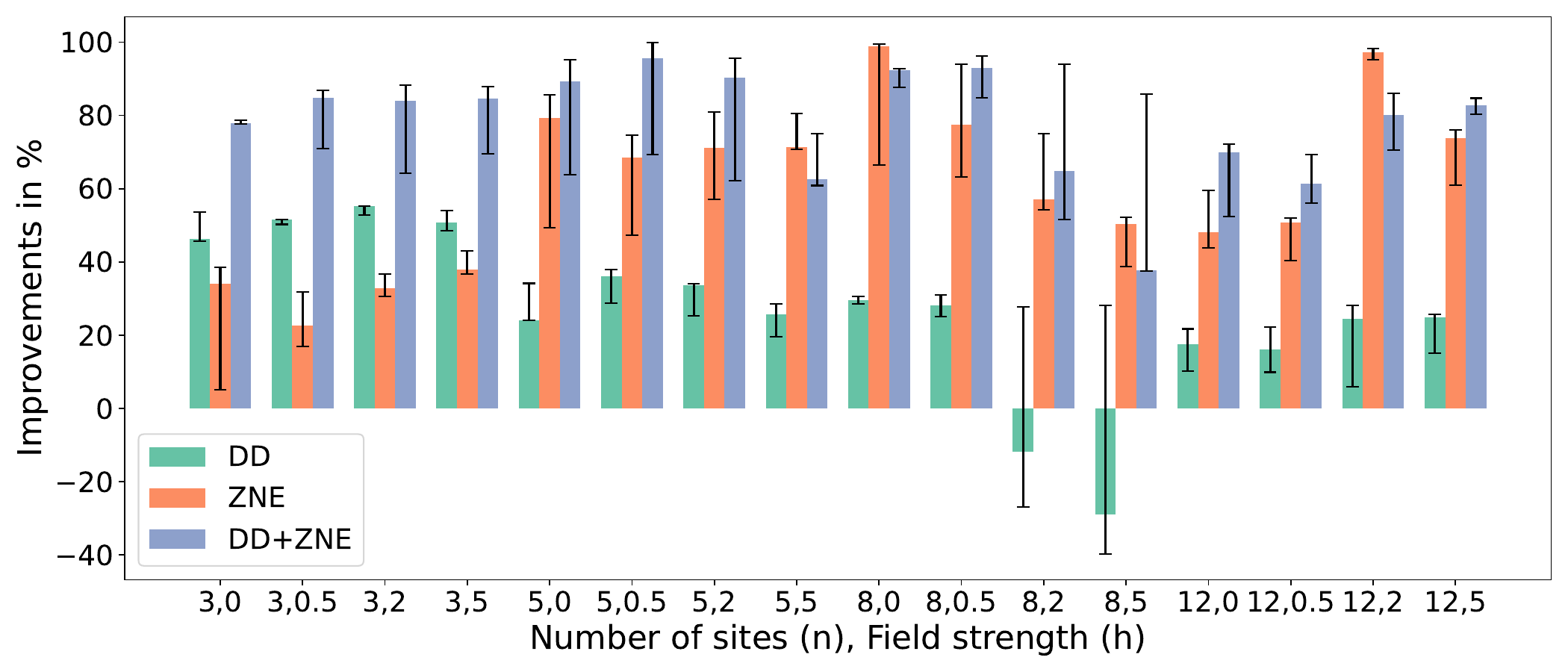}
    \caption{Percentage improvements in energy gap for Ising model ground state estimation.}
    \label{fig:ising_gsp}
\end{figure*}

\begin{figure*}
    \subfigure[]{
        \includegraphics[width=0.40\textwidth]{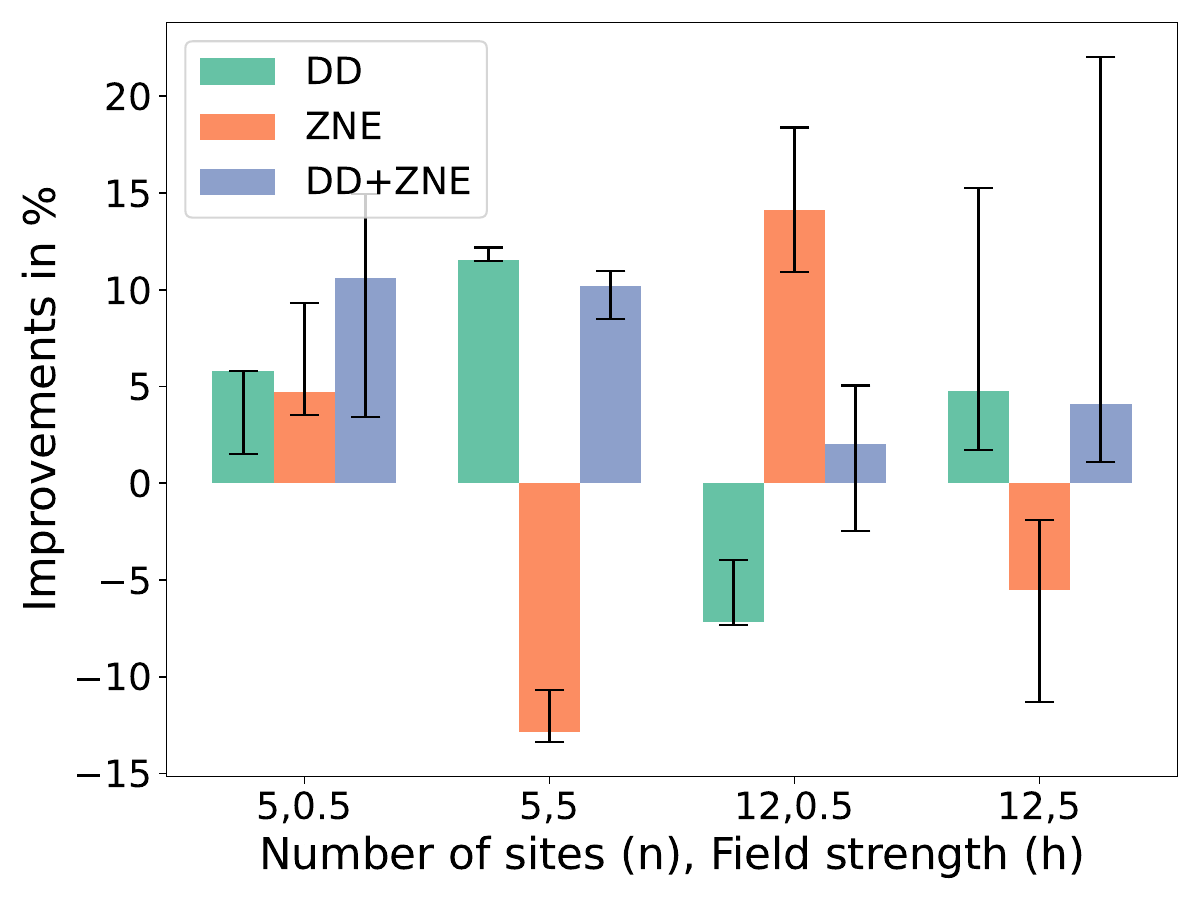}
        \label{fig:heisenberg_evo}
    }
    \hspace{30pt}
    \subfigure[]{
        \includegraphics[width=0.40\textwidth]{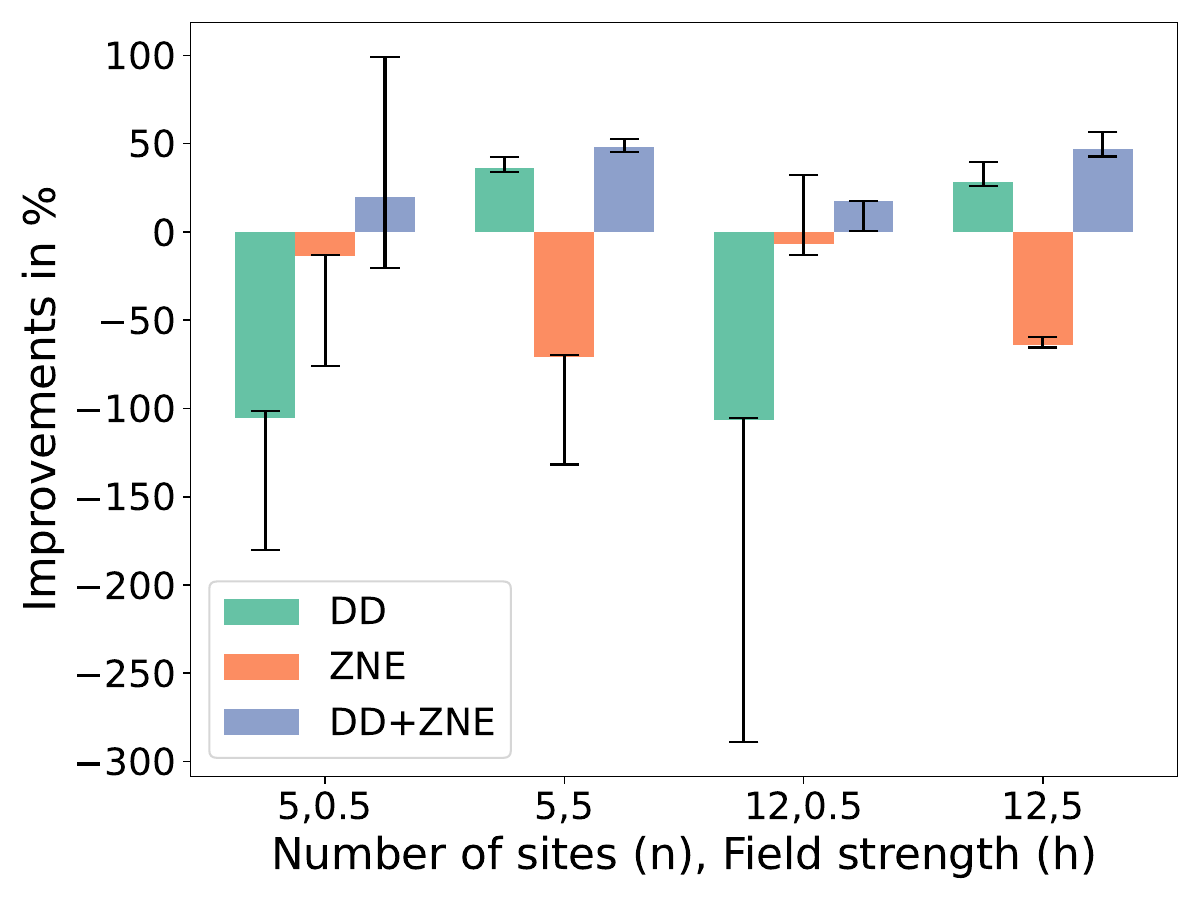}
        \label{fig:ising_evo}
    }
    \caption{
     Percentage improvements in the energy difference after time evolution for
    Heisenberg model ~\ref{fig:heisenberg_evo} and Ising model ~\ref{fig:ising_evo}.
    }
\end{figure*}

The experimental results across both ground state preparation and time evolution demonstrate that while individual mitigation techniques vary in efficacy depending on the system size ($n$) and field strength ($h$), the combined DD+ZNE approach consistently yields the highest improvements.
In Figures ~\ref{fig:heisenberg_gsp}, ~\ref{fig:ising_gsp}, ~\ref{fig:heisenberg_evo} and ~\ref{fig:ising_evo} the experiments laid out in Section ~\ref{sec:experimental_setting} are repeated three times and the median value of the improvement percentage is reported.

\subsection{Ground state estimation}

In the ground state experiments, whose results are shown in Figures ~\ref{fig:heisenberg_gsp} and ~\ref{fig:ising_gsp} we observe:

\begin{itemize}
    \item For both the TFIM and Heisenberg models, the DD+ZNE configuration (blue bars) consistently outperforms individual strategies, often achieving over 60\% reduction in the energy gap across most system sizes.
    \item As the number of sites ($n$) increases, the improvement from DD-only (green bars) tends to decrease, and may sometimes even increase the energy gap. This suggests that as the circuit grows, the accumulation of crosstalk and multi-qubit dephasing during MCM/FF latencies becomes more challenging to suppress with DD alone.
\end{itemize}

\subsection{Time evolution}

The time evolution results in Figures ~\ref{fig:heisenberg_evo} and ~\ref{fig:ising_evo} reveal a more complex noise profile :
\begin{itemize}
    \item Individual techniques (DD or ZNE) occasionally show negative improvement (increase in energy gap). This is likely due to the high variance of the Trotterized energy expectation value arising from the increased gate depth and the stochastic nature of MCM/FF errors.
    \item Despite the volatility of individual methods, the DD+ZNE strategy remains robust, maintaining positive improvement levels. In TFIM with $h=5$, the combined method recovers 15\%-99\% of the energy gap, whereas in the more complex Heisenberg model, improvements are more modest (with median performance reaching ~10\% and the maximum reaching ~20\%).
    \item The Heisenberg model simulations show significantly lower overall improvement percentages compared to the TFIM model. This highlights the increased difficulty of mitigating errors in non-commuting dynamics where $R_{XX}$ and $R_{YY}$ terms introduce additional depth overhead and decoherence.
\end{itemize}

In summary, the combined DD+ZNE strategy works best in mitigating the noise introduced by MCM and FF operations.

\section{Conclusion}
\label{sec:conclusion}

In conclusion, our study validates that dynamic circuits can be successfully deployed for complex many-body simulations provided they are supported by an integrated error-mitigation framework. By leveraging DD to mitigate decoherence during classical processing latencies and ZNE to mitigate systematic gate errors, we have shown that dynamic circuits can maintain relatively high fidelity even as system sizes increase. For future work, we plan to incorporate probabilistic error cancellation (PEC) ~\cite{gupta2024probabilistic} for MCMs and probabilistic readout error mitigation (PROM) ~\cite{koh2026readout} to further refine the accuracy of dynamic primitives. Moreover, we aim to extend these mitigation strategies to quantum error-correction primitives such as syndrome-extraction circuits, tightening the bridge between near-term dynamic circuits and the eventual transition to fault-tolerant operation.

\section*{Code availability}
Code used to perform the experiments and analyze the results is publicly available online: 
\href{https://github.com/sumeetshirgure/dynemsim}{https://github.com/sumeetshirgure/dynemsim}.

\begin{acks}
S.S. is supported by the University of Central Florida ORCGS Doctoral
Fellowship award.
This research used resources of the National Energy Research
Scientific Computing Center, a DOE Office of Science User Facility
supported by the Office of Science of the U.S. Department of Energy
under Contract No. DE-AC02-05CH11231 using NERSC award
NERSC DDR-ERCAP0038372.
\end{acks}

\bibliographystyle{ACM-Reference-Format}
\bibliography{main}

\end{document}